\documentclass[runningheads,orivec]{llncs}

\usepackage[T1]{fontenc}
\usepackage{graphicx}
\usepackage[bookmarks,bookmarksopen]{hyperref}
\usepackage{amsmath}
\usepackage{amssymb}
\usepackage{color}
\usepackage{paralist} 
\usepackage{stmaryrd} 

\usepackage{tikz}
\usetikzlibrary{arrows.meta, positioning}

\tikzset{
  ew/.style={
    to path={(\tikztostart.east) -- (\tikztotarget.west)}
  }
}

\usepackage{local}

\usepackage{todonotes}

\newcommand{\smalltt}[1]{\small \texttt{#1}}

\begin{document}

\title{The Dependently Typed Higher-Order Form \\ for the TPTP World}

\titlerunning{DTF: The TPTP DHOL Form}

\author{Daniel Ranalter\inst{1}\orcidID{0009-0006-2861-548X} \and
Cezary Kaliszyk\inst{2,1}\orcidID{0000-0002-8273-6059} \and
Florian Rabe\inst{3}\orcidID{0000-0003-3040-3655} \and
Geoff Sutcliffe\inst{4}\orcidID{0000-0001-9120-3927}}

\authorrunning{D. Ranalter et al.}

\institute{University of Innsbruck, Computational Logic, Austria\\
\email{d.ranalter@gmail.com}
\and
University of Melbourne, School of Computing, Australia\\
\email{ckaliszyk@unimelb.edu.au}
\and
University of Erlangen-Nuremberg, Computer Science, Germany\\
\email{florian.rabe@fau.de}
\and
University of Miami, Department of Computer Science, USA\\
\email{geoff@cs.miami.edu}}

\maketitle

\begin{abstract}
  Much of the current research and development in the field of automated reasoning builds on the
  infrastructure provided by the TPTP World.
  The TPTP language for logical formulae is central to the far-reaching adoption of the TPTP World. 
  This paper introduces the Dependently Typed higher-order Form (DTF) of the TPTP language. 
  It takes advantage of already established binders in the syntax, and is thus a minimally 
  intrusive extension to the Typed Higher-order Form (THF). 
  A starting set of over 100 problems is provided to exhibit the usefulness and incite interest 
  in DTF. 
  Some tools that are already able to reason about problems in the DTF language are discussed. 
  
\end{abstract}

\section{Introduction}
The TPTP World~\cite{S24} is a well-established infrastructure that supports research, 
development, and deployment of Automated Theorem Proving (ATP) systems. 
The TPTP language~\cite{Sut23-IGPL} is one of the keys to the success of the TPTP World.
It has variants that support uniform expression of logical formulae across a wide range of logics. 
The TPTP language is used for writing both problems and solutions, which enables convenient 
communication between ATP systems and tools.
The majority of modern ATP systems accept input in TPTP syntax.
The TPTP language variants that form the basis for this work are the monomorphic and polymorphic
typed higher-order forms (TH0 and TH1)~\cite{SB10,KSR16} (see Section~\ref{TPTPWorld} for the
background and further variants).

All the existing typed TPTP language variants are \emph{simply} typed. 
However, there is a steady increase of interest in \emph{dependently} typed systems, such as 
Agda~\cite{BDN09}, Rocq~\cite{BertotCasteran2004,Rocq}, and Lean~\cite{dMU21}.
This interest extends to the SMT community, where the proposed version 3.0 of SMT-LIB is to
include dependent types\footnote{%
\href{https://smt-lib.org/version3.shtml}{smt-lib.org/version3}}.
Dependent types allow for the elegant formulation of complex data structures, possibly even 
a direct encoding of correctness properties. 
This paper introduces the Dependently Typed higher-order Form (DTF) of the TPTP language. 

While dependent types are frequently used in interactive theorem proving, Automated Theorem 
Proving (ATP) has yet to embrace dependent types.
Rothgang et al. made first steps towards bringing ATP and dependent types together, by introducing 
dependently typed higher-order logic (DHOL)~\cite{RRB23,RRB23ext}.
With only two minor extensions to the familiar syntax of Church-style HOL~\cite{C40}, DHOL 
makes dependent types easily accessible: HOL base types are extended into dependent base types 
that can take term arguments, and the function type \(A \sTo B\) is changed into a dependent 
function type \(\dTo xAB\).
Originally DHOL did not allow quantifying over types or stating the equality of types, but a 
polymorphic version is in development.

As in FOL and HOL, DHOL allows arbitrary axioms that may constrain equality of terms in 
undecidable ways, and consequently DHOL's type checking is undecidable 
(see Section~\ref{ssec:tcheck}).
To manage this complication Rothgang et al. provide an algorithm that reduces the well-formedness 
of a statement to a set of proof obligations.
Thus theorem proving is needed to check the well-formedness of a problem's formulae, not 
just to prove the conjecture.
Happily, typically that does not make it harder to prove the conjecture.
To increase ATP support for DHOL, Rothgang et al. define a translation from well-typed DHOL to HOL 
that preserves provability in both directions, thereby making DHOL available for regular HOL ATP 
systems, albeit without leveraging DHOL's dependent types for more efficient proving.
Furthermore, the translation introduces additional axioms capturing the constraints of the
dependent types, thereby potentially complicating proof search.
Several interactive theorem provers had previously employed the same idea, sacrificing decidable 
typing to gain the expressivity of dependent types, while keeping the general feel of the language 
simple.
Most importantly, PVS~\cite{OS97} essentially contains DHOL as a fragment, but extends it 
beyond the capabilities of current \emph{automated} provers.
Mizar~\cite{TB85}, using soft typing on top of first-order set theory, can also capture DHOL-like 
features.

A detail missing from the original formulation of DHOL was the choice operator. Ranalter et
al. investigated the effects of losing the non-emptiness constraint in DHOL on Hilbert's choice
in \cite{RBK24}. To this end, they extended the -- to the authors knowledge -- first native
implementation of DHOL into the ATP system Lash, by Niederhauser et al.~\cite{NBK24}. 
Their experiments strongly suggest that native reasoning in DHOL significantly outperforms 
reasoning on translated problems.

This work describes how DHOL is being integrated into the TPTP World, in a new TPTP language
variant ``Dependently Typed higher-order Form'' (DTF), with monomorphic and polymorphic 
subvariants (DT0 and DT1).
DTF requires only very minor changes to the familiar TPTP language syntax, mostly using 
existing notions for binders and application operators, thereby providing the ATP community with 
the necessary foundations on which research into dependently typed automated reasoning can thrive.
A set of over 100 problems in DTF, taken from several different sources, has been curated as an 
initial contribution to the TPTP problem library.
The problems provide a spread of interesting formulations focusing on a variety of difficulty 
levels in proving the conjecture as well as in type checking. 

Section~\ref{s:prelims} reviews the TPTP World and establishes the necessary background for
DHOL, slightly generalizing the original DHOL definition to make it more suitable for TPTP. 
Section~\ref{s:thd} introduces the new DTF form.
Section~\ref{s:probs} gives a short overview of the starting set of problems, and 
Section~\ref{s:impl} introduces tools that already support the new form.
Finally, Section~\ref{s:conc} concludes and gives an outlook over future work.

\section{Preliminaries}\label{s:prelims}
\subsection{The TPTP World and Infrastructure}\label{ss:tptp}
\label{TPTPWorld}

The TPTP World infrastructure includes
the TPTP language~\cite{SS+06},
the TPTP problem library~\cite{Sut09},
the TSTP solution library~\cite{Sut10},
the SZS ontologies~\cite{Sut08-KEAPPA},
the Specialist Problem Classes (SPCs) and problem difficulty ratings~\cite{SS01},
SystemOnTPTP~\cite{Sut00-CADE-17} and StarExec~\cite{SST14},
and the CADE ATP System Competition (CASC)~\cite{S16}.
The problem library is a
large collection of Thousands of Problems for Theorem Proving -- hence the name. 
The problem library release v9.1.0 contains over 26000 problems from over 50 different domains, 
written in the TPTP language.
The problems are categorized into Specialist Problem Classes according to their syntactic and
logical status.
The TSTP solution library is the result of running numerous ATP systems on the problems in that 
library and collecting their output. 
The TPTP and TSTP libraries provide the basis for assigning a difficulty rating to each problem, 
according to which ATP systems are able to solve the problem.

The most salient feature of the TPTP World for this work is the TPTP language.
Originally the TPTP language supported only first-order clause normal form (CNF)~\cite{SS98}.
Over time, more complex logics were added, starting with first-order form (FOF) in TPTP release 
v2.0.0~\cite{Sut09}. 
Releases v3.0.0 and v4.0.0 added monomorphic typed higher-order (TH0)~\cite{SB10} and
monomorphic typed first-order (TF0)~\cite{S+12} forms to the mix respectively. These got
extended to their polymorphic variants TF1 and TH1 in releases v5.0.0~\cite{BP13} and
v6.0.0~\cite{KSR16}. Release v7.0.0 of the TPTP started to include extended typed
first-order form (TXF)~\cite{SK18} which extends the typed first-order form with conditionals,
let expressions, and boolean terms. 
All the listed extensions to the TPTP are classical in nature. 
This changed with the addition of non-classical typed first-order form (NTF) in release 
v9.0.0~\cite{S+22}.
A general principle of the TPTP language is: ``We provide the syntax, you provide the semantics''.
As such, there is no a priori commitment to any semantics for each of the language forms, 
although in almost all cases the intended logic and semantics are well known.

Problems and solutions are built from {\em annotated formulae} of the form
\begin{center}
{\em language}{\tt (}{\em name}{\tt ,}
{\em role}{\tt ,}
{\em formula}{\tt ,}
{\em source}{\tt ,}
{\em useful\_info}{\tt )}
\end{center}
The {\em language}s supported are {\smalltt{cnf}} (clause normal form), {\smalltt{fof}}
(first-order form), {\smalltt{tff}} (typed first-order form), and {\smalltt{thf}}
(typed higher-order form).
The {\em role}, e.g., {\smalltt{axiom}}, {\smalltt{lemma}}, {\smalltt{conjecture}}, defines the 
use of the formula.
In a {\em formula}, terms and atoms follow Prolog conventions -- functions and predicates start 
with a lowercase letter or are {\tt '}single quoted{\tt '}, and variables start with an uppercase 
letter.
The language also supports interpreted symbols that either start with a {\tt \$}, e.g., the 
truth constants {\smalltt{\$true}} and {\smalltt{\$false}}, or are composed of 
non-alphabetic characters, e.g., integer/rational/real numbers such as 27, 43/92, -99.66.
The logical connectives in the TPTP language are
{\tt !>}, {\tt ?*}, {\tt @+}, {\tt @-}, {\tt !}, {\tt ?}, {\tt {\raisebox{0.4ex}{\texttildelow}}}, 
{\tt |}, {\tt \&}, {\tt =>}, {\tt <=}, {\tt <=>}, and {\tt <{\raisebox{0.4ex}{\texttildelow}}>},
for the mathematical connectives
$\Pi$, $\Sigma$, choice (indefinite description), definite description,
$\forall$, $\exists$, $\neg$, $\vee$, $\wedge$, $\Rightarrow$, $\Leftarrow$, $\Leftrightarrow$, 
and $\oplus$ respectively.
Equality and inequality are expressed as the infix operators {\tt =} and {\tt !=}.
The {\em source} and {\em useful\_info} are optional.

\subsection{Dependently Typed Higher-Order Logic}

Dependently typed higher-order logic (DHOL) is an extension of Church's higher-order logic
(HOL)~\cite{C40} introduced by Rothgang et al.~\cite{RRB23}. It takes the widely supported HOL
and equips it with dependent types, i.e., types that take term arguments. As such, it is a
classical and extensional type theory, as opposed to the theory used in 
Rocq~\cite{BertotCasteran2004,Rocq}, Lean~\cite{dMU21}, or others~\cite{BDN09,PS99} 
that rely on an intensional type theory. 
Notable exceptions to this trend are PVS~\cite{ROS98}, NuPRL~\cite{nuprl}, and F*~\cite{S+16}.

The extensionality of DHOL comes at the cost of making type checking undecidable because it
must consider term equality, which may be subject to arbitrary axioms.
Essentially, typing becomes undecidable if a type depends on a type for which equality is 
undecidable.
This is because type checking \(t\) against type \(a\ n\) must be done by inferring the type of $t$, say \(a\ m\), and then checking \(a\ m = a\ n\), and thus
\(m = n\).
If all dependent type symbols depend only on types for which equality is decidable (e.g., the examples below where we only use natural numbers with Presburger arithmetic), type checking is decidable.
Otherwise, e.g., when using types depending on natural numbers with Peano arithmetic, type checking is undecidable.

The gain of having judgmental and provable equality coincide is significant:
It positions DHOL much closer to how mathematics is usually done in the context of ATP.
The availability of dependent types allows the elegant definition of data structures such as lists 
of fixed-length, intervals of numbers, or vector spaces over some field.
It also allows encoding constraints in the types, which can remove the need for lengthy and error-prone 
guards in programming and track invariants useful for theorem proving.
The cost -- which might seem steep at first glance -- is mitigated by the ever-increasing 
performance of ATP systems, and the fact that in many cases the proof obligations resulting from 
type checking are much simpler than the original proving problem.

The changes to the TPTP syntax to accommodate DTF are small: the definition of the simple base 
type is changed to a type that can accept term arguments, and the simple function type 
\(A \sTo B\) is changed to \(\dTo xAB\). 
This makes it possible to let the result type of the function depend on the specific term of the 
argument.

Figure~\ref{fig:grammar} gives the grammar of DHOL. 
A dependent base type \(a\) with arity \(n\) is written \(a:\Pi x_1:A_1,\cdots,x_n:A_n.\tp\), and 
it is a \emph{simple} base type if \(n = 0\). Declarations of this form are part of the theory
against which the type checking procedure is performed. 
In addition to base type declarations, theories may declare constant symbols \(c\) and axioms
\(\ass F\).
A context specifies typed variables and assumptions.
Contexts are superficially similar to theories, but denote \emph{local} declarations, and as such, 
do not contain type declarations.
\(\circ\) and \(\bullet\) denote the empty theory and context respectively. 
The order in a theory or context matters because the well-typedness of declarations 
might depend on preceding axioms.
Types, as they appear in statements and typing judgements, are either fully applied base types,
(dependent) function types, or classical booleans \(o\).
Terms are built from variables/constants, lambda abstraction, application, and the usual 
connectives and quantifiers.
Regular HOL can be recovered by omitting the \hili{highlighted} elements -- this is exactly the 
case when the arity of all base types is 0. 

\begin{figure}[h]
  \centering
  \begin{tabular}{l c l r}
    \(T, U\) & ::= & \(\circ\) | \(T,\; a: \hili{(\Pi x:A.)^*} \tp\)
                     | \(T,\;c:A\) | \(T,\;\ass F\) & \quad theories \\
    \(\Gamma, \Delta\) & ::= & \(\bullet\) | \(\Gamma,\, x:A\) | \(\Gamma,\, \ass F\) & context \\
    \(A,B\) & ::= & \(a\,\hili{t_1\,\cdots\,t_n}\) | \(\hili{\Pi x:}A.B\) | \(o\) & types \\
    \(t,u,F,G\) & ::= & \multicolumn{2}{l}{\(x\) | \(c\) | \(\lambda x:A.t\) | \(t\,u\) | \(\forall x:A.F\) | \(\exists x:A.F\) | \(F \Rightarrow G\)} \\
    & | & \(F \land G\) | \(F \lor G\) | \(\bot\) | \(\top\) | \(\neg F\) | \(t =_A u\) & terms (incl. formulae $F,G$)\\
  \end{tabular}  
  \caption{The grammar of DHOL}
  \label{fig:grammar}
\end{figure}

The following example encodes the familiar notion of fixed-length lists. 
As prerequisites, we give the usual notion of natural numbers in a simple type $\nat$ and a simple type $\chars$ of 
characters for the elements of the lists:
\begin{gather*}
  \nat :\tp
  \qquad 0:\nat
  \qquad \suc:\nat\sTo\nat
  \qquad\+:\nat\sTo\nat\sTo\nat\\
  \ass\forall n:\nat.\+\; 0\; n =_{\nat} n
  \qquad \ass\forall n, m:\nat.\+\; (\suc\; n)\; m=_{nat} \suc\; (\+\; n\; m)\\
  \chars :\tp \qquad \aChar:\chars \qquad \bChar:\chars \qquad ...\\
\end{gather*}
Then $\vect\,n$ encodes the type of fixed-length lists of characters of length $n$:
\begin{gather*}
  \vect:\Pi n:\nat.\tp
  \quad \nil:\vect\; 0
  \quad \cons:\dTo{x}{\nat}{\chars \sTo \vect\; n \sTo \vect\; (\suc\ n)}\\
  \app:\dTo{n,m}{\nat}{\vect\; n \sTo \vect\; m \sTo \vect\; (\+\; n\; m)}
\end{gather*}

\paragraph{Dependent Connectives}
In DHOL it is desirable to make the binary connectives conjunction, implication, and
disjunction dependent in the sense that the well-formedness of the second argument may assume
the truth (for conjunction and implication) or the falsity (for disjunction) of the first argument.
Consider the statement \(a =_A b \impl f\: a =_{B(a)} f\: b\). 
The well-formedness of the right-hand side requires the left-hand side as a premise.
More precisely, $\Gamma\vdash F:\bool$ resp. $\Gamma\vdash F$ expresses that $F$ is a 
well-formed resp. provable formula in context $\Gamma$.
The definition of well-formed formulae is:
\begin{center}
\begin{tabular}{l@{\quad if\quad}l}
$\Gamma\vdash F\impl G$ & $\Gamma\vdash F$ and $\Gamma,\,\hili{\ass F}\vdash G$\\
$\Gamma\vdash F\wedge G$ & $\Gamma\vdash F$ and $\Gamma,\,\hili{\ass F}\vdash G$\\
$\Gamma\vdash F\vee G$ & $\Gamma\vdash F$ and $\Gamma,\,\hili{\ass \neg F}\vdash G$
\end{tabular}
\end{center}
where the \hili{marked} parts make the connectives dependent.
The usual natural deduction proof rules of implication and conjunction are the same as for the 
non-dependent versions.
The proof rules for disjunction are adjusted as follows:
\[\rul{\Gamma\vdash F\quad \Gamma,\,\hili{\ass\neg F}\vdash G:\bool}{\Gamma\vdash F\vee G} \quad
  \rul{\Gamma,\,\hili{\ass \neg F}\vdash G}{\Gamma\vdash F\vee G} \quad
  \rul{\Gamma,\,\ass F\vdash C\quad \Gamma,\,\hili{\ass \neg F},\,\ass G\vdash C}{\Gamma,\,\ass F\vee G \vdash C}
\]

As usual, it is possible to choose some connectives as primitives, from which the others are 
defined.
Rothgang et al. choose equality and implication.
Contrary to HOL, they included implication because they could not define the dependent binary 
connectives solely from equality.
For the TPTP World, it is better not to choose primitive connectives -- that choice should be left 
to the ATP system developers.
Therefore DTF extends the work by Rothgang et al. to make all connectives primitive.
ATP systems can choose which connectives to treat as abbreviations, but in doing so must take the 
dependent nature of the connectives into account.
Note that dependent connectives break the commutativity of conjunction and disjunction.
While seemingly disruptive, sacrificing commutativity in this way is common practice, e.g., for 
short-circuit evaluation of Boolean terms in programming languages.
To clarify the impact on theorem proving, Table~\ref{proofrules} summarizes typical proof rules 
for FOL and their status in DHOL.
Roughly speaking, all rules that do not affect the order of subformulae remain sound, while the
rest of the rules require the additional check to ensure the result remains well-formed.
In particular, all rules needed to perform CNF or clause normal form transformations remain 
available.

\begin{table}[h]
  \centering
  \caption{Typical proof rules for FOL and their status in DHOL}
\begin{tabular}{|l|l|}
\hline
\textbf{Rule} & \textbf{Holds in DHOL} \\
\hline
\multicolumn{2}{|c|}{For disjunction and conjunction}\\
associativity & \checkmark \\
commutativity & Only if both sides are well-formed \\
idempotence, e.g., $A\wedge X\wedge A \Leftrightarrow A\wedge X$ & \checkmark (Drop the \emph{second} occurrence)\\
de Morgan laws & \checkmark \\
distributivity of one over the other & \checkmark \\
absorption, e.g., $A\wedge (A\vee B) \Leftrightarrow A$ & \checkmark \\
\hline
\multicolumn{2}{|c|}{For implication}\\
$A\impl B \Leftrightarrow \neg A\vee B$ & \checkmark\\
$\neg(A\impl B) \Leftrightarrow A\wedge\neg B$ & \checkmark\\
$\neg (A\impl B) \Leftrightarrow \neg B\impl \neg A$ & Only if both sides are well-formed\\
\hline
\multicolumn{2}{|c|}{For quantifiers and equality}\\
all rules & \checkmark\\
\hline
\multicolumn{2}{|c|}{Common calculus rules}\\
classical reasoning & \checkmark \\
weakening & \checkmark \\
contraction &  \checkmark (Drop the \emph{second} occurrence)\\
exchange & Only if still well-formed\\
cut & \checkmark \\
resolution & Only if the clauses remain well-formed\\
\hline
\end{tabular}
\label{proofrules}
\end{table}

Developing advanced calculi for DHOL is beyond the scope of this paper.
However, for example, one way to generalize resolution is to store clauses as lists 
$[L_1,\ldots,L_n]$ where the well-formedness of each $L_i$ may depend on $\neg L_j$ for $j<i$. 
Resolving $[A,\vec{L}]$ and $[\neg A,\vec{M}]$ to $[\vec{L},\vec{M}]$ is sound if the resolvent 
is well-formed, i.e., if the well-formedness of the $L_i$ resp. $M_i$ does not depend on 
$\neg A$ resp. $A$.

\subsubsection{Polymorphic DHOL.}
\label{ss:poly}

DHOL as presented in the previous section and \cite{RRB23} is monomorphic.
ATP for polymorphic DHOL, as well as proofs of properties for such an extension of the calculus, 
is ongoing parallel work.
Polymorphic logics are already available in the TPTP language, so it is natural to offer
polymorphic DTF.
All the polymorphic example problems considered so far use only shallow/rank-1 polymorphism in line with the existing polymorphic first- and higher-order forms for TPTP.


\subsubsection{Choice.}

Hilbert's choice operator has been part of HOL since its inception by Church~\cite{C40}. 
As such, it is natural to include it in DHOL. 
This introduces some complications: Due to the usual non-emptiness constraint on types, the 
semantics of choice are clear in HOL. 
However, DHOL no longer abides by this constraint, requiring a design decision that affects 
well-typedness and provability.
Experiments done in \cite{RBK24} suggest that the variant of choice dubbed ``strong choice'' 
results in more efficient automated reasoning. 
The eponymous characteristic of strong choice is the requirement that \(\exists x:A.t\) needs 
to be true for \((\varepsilon x:A.t):A\) to be well-typed.
Such a requirement for typing fits well with DHOL in general, and as ATP is the main concern this 
is the variant of choice, as it were. 
The problem set described in Section~\ref{s:probs} includes some examples supporting this variant.

\subsubsection{Translation.}
\label{ss:erasure}

In order to take advantage of the ATP systems available for regular HOL, Rothgang et al. define 
a dependency-erasure~\cite{RRB23}, and thereby a translation from DHOL into regular HOL. 
They also prove that this translation is sound and complete for well-typed DHOL problems. 
Due to this result, and the implementation of the translation into the preprocessor of the 
Leo-III theorem prover~\cite{SB21}, there existed reasoning support for DHOL even before native 
DHOL reasoning was implemented in the Lash ATP system by Niederhauser et al.~\cite{NBK24}.
Information lost due to the erasure of term dependencies is captured in Partial Equivalence
Relations (PERs) -- symmetric and transitive relations on pairs of terms -- with the idea
that the relation is reflexive exactly for those terms that were previously of the same
dependent type.
The translation is shown in Figure~\ref{fig:trans}.
The translation \(\tr{t}\) of a term \(t\) is defined inductively on the structure of the terms.
The erasure of one type declaration results in three erased declarations: the erased type, the
PER constant and an axioms stating it's properties.
The definition of the erasure on \(\forall\)- and \(\exists\)-quantified terms is notable as it
uses a PER as guard on the argument.
To see why, note that, e.g., \(\forall x:A.t\) can be defined in terms of equality as \(\lambda
x:A. t =_{A\rightarrow o} \lambda x:A.\top\). 
The erasure creates a PER from this typed equality with the guarded input in the premise, and the 
erased term in the consequence of the implication as seen in the erasure of \(\forall\).

\begin{figure}[tphb]
\begin{align*}
\multispan{2}{theories\hfil}& \multispan{2}{contexts}\\
\overline{\circ} \:&=\: \circ & \overline{\bullet} \:&=\: \bullet \\
\overline{T, D} \:&=\: \overline{T}, \overline{D} & \overline{\Gamma, D} \:&=\: \overline{\Gamma}, \overline{D} \\
\overline{c: A} \:&=\: c: \overline{A},\; \ass\per{A}\: c\: c & \overline{x: A} \:&=\: x: \overline{A},\; \ass\per{A}\: x\: x\\
\overline{\ass F} \:&=\: \ass \overline{F} & \overline{\ass F} \:&=\: \ass \overline{F}\\
\multispan{4}{$\overline{a:\Pi x_1:A_1.\: \cdots \: \Pi x_n:A_n.\:\tp}\:=$\hfil} \\
 & \multispan{3}{$a\: \tp$\hfil} \\
 & \multispan{3}{$\per{a}: \overline{A_1} \to \cdots \to \overline{A_n} \to a \to a \to o$\hfil} \\
 & \multispan{3}{$\ass\forall  x_1:\overline{A_1}.\: \cdots \forall x_n:\overline{A_n}.\: \forall u,v:
   a.\: \per{a}\ x_1\:\cdots\:x_n\ u\ v \impl u =_a v$\hfil} \\[.2cm]
\multispan{4}{terms and types\hfil} \\
\overline{c} \:&=\: c & \overline{x} \:&=\: x \\
\overline{o} \:&=\: o & \overline{a\ t_1\ \ldots\ t_n} \:&=\: a \\
\overline{\dTo xAB} \:&=\: \overline{A} \sTo \overline{B} & \overline{\lambda x : A.t} \:&=\: \lambda x:\overline{A}.\:\overline{t} \\
\overline{\neg t} \:&=\: \neg \overline{t} & \overline{t\: u} \:&=\: \overline{t}\: \overline{u} \\
\overline{t \impl u} \:&=\: \overline{t} \impl \overline{u} &
\overline{t =_{A} u} \:&=\: \per{A}\: \overline{t}\: \overline{u} \\
\overline{t \land u} \:&=\: \overline{t}\land \overline{u} &
\overline{t \lor u} \:&=\: \overline{t} \lor \overline{u}\\
\overline{\bot} \:&=\: \bot & \overline{\top} \:&=\: \top\\
\overline{\forall x: A.t} \:&=\: \forall x:\overline{A}.\: \per{A}\: x\: x \impl \overline{t} &
\overline{\exists x: A.t} \:&=\: \exists x:\overline{A}.\: \per{A}\: x\: x \land \overline{t} \\
\end{align*}
PER for each type
\begin{align*}
 \per{o}\: t\: u \:&=\: t =_o u \\
 \per{(a\ t_1\ \ldots\ t_n)}\: u\: v \:&=\: \per{a}\ \overline{t_1}\: \cdots \: \overline{t_n}\ u\ v \\
 \per{(\dTo xAB)}\: t\: u\:&=\: \forall x,y:\overline{A}.\: \per{A}\: x\: y \impl \per{B} (t\: x) (u\: y)\\
\end{align*}
\caption{The translation from DHOL to HOL}
\label{fig:trans}
\end{figure}

As an example of erasure, consider the list of \texttt{char}s [a, b], represented by a
term \(\cons\; 1 \; \aChar \; (\cons\; 0\; \bChar\; \nil) \) of type \(\vect\; 2\), where 
\(0, \suc\; 0, \suc\; (\suc\; 0), ...\) is abbreviated as \(0, 1, 2, ... \). 
Applying the erasure gives \(\cons\; \aChar\; (\cons\; \bChar\; \nil)\) of type \(\vect\).
A predicate would be generated, establishing that this particular list is in the PER of vectors 
of length 2: \(\per{\vect}\; 2\; t\; t\) where t stands for \(\cons\; \aChar\;
(\cons\; \bChar\; \nil)\). 
While one might think that unary predicates would be sufficient as a type 
guard, PERs becomes necessary to express the typing and equality of higher-order functions: functions are well-typed if they map well-typed inputs to well-typed outputs, and they are equal if they agree on well-typed inputs.

\section{DTF}\label{s:thd}
After establishing the theoretic background, this section presents the realization of DHOL in
the TPTP language. Syntax and semantics are given, as well as an exposition to the problem of
type checking.

\subsection{Syntax}

The syntax of DTF requires almost no change to the existing TPTP syntax. 
The TPTP language already defines the \texttt{!>} binder for types. 
In the typed TPTP language variants it is currently used for only polymorphism, e.g.,
\begin{center}
\texttt{cons : !>[A: \$tType]: ( A > ( list @ A ) > ( list @ A ) )}
\end{center}
\noindent
is a type declaration for a polymorphic \texttt{cons}.
The TPTP syntax does not forbid listing terms in the types of such variable lists. 
This fact is used to unobtrusively extend TPTP by dependent types. 
A dependent type symbol declaration is written with \(m\) terms of \(n\) types as 
\begin{center}
\texttt{a : !>[\(x_1:A_1\), ..., \(x_m:A_n\)]:\$tType}
\end{center}
\noindent
or alternatively 
\begin{center}
\texttt{a : \(A_1\) > ... > \(A_n\) > \$tType}. 
\end{center}
\noindent
Such types use the application operator \texttt{@}, to instantiate the terms to the dependent 
type: 
\begin{center}
\texttt{a @ \(t_1\) @ ... @ \(t_m\)}. 
\end{center}
\noindent
In polymorphic problems, the variable list is prepended with the type variables, which 
may appear in the same binder.
An example of a problem in DTF is shown in Figure~\ref{fig:syn}.

\begin{figure}
\begin{verbatim}
thf(elem_type,type,    elem: $tType ).
thf(nat_type,type,     nat: $tType ).
thf(zero_type,type,    zero: nat ).
thf(suc_type,type,     suc: nat > nat ).
thf(plus_type,type,    plus: nat > nat > nat ).
thf(list_type,type,    list: nat > $tType ).
thf(nil_type,type,     nil: list @ zero ).
thf(cons_type,type,    cons: 
    !>[N: nat] : (elem > (list @ N) > (list @ (suc @ N))) ).
thf(app_type,type,     app: 
    !>[N: nat,M: nat] : ((list @ N) > (list @ M) > 
                         (list @ (plus @ N @ M))) ).

thf(ax1,axiom,
    ! [N: nat] : ((plus @ zero @ N) = N) ).

thf(ax2,axiom,
    ! [N: nat,X: list @ N] : ((app @ zero @ N @ nil @ X) = X) ).

thf(plus_assoc,axiom,
    ! [M1: nat,M2: nat,M3: nat] :
      ( (plus @ M1 @ (plus @ M2 @ M3))
      = (plus @ ( plus @ M1 @ M2) @ M3)) ).

thf(list_app_assoc_base,conjecture,
    ! [M2: nat,L2: list @ M2,M3: nat,L3: list @ M3] :
      ( (app @ zero @ (plus @ M2 @ M3) @ nil @ 
          (app @ M2 @ M3 @ L2 @ L3))
      = (app @ (plus @ zero @ M2) @ M3 @ 
          (app @ zero @ M2 @ nil @ L2) @ L3)) ).
\end{verbatim}
 \caption{\label{fig:syn}The base case of associativity of append on fixed-length lists.}
\end{figure}

\subsection{Type Checking}
\label{ssec:tcheck}

Due to equality reflection, type checking for DHOL is, in general, undecidable. 
Nevertheless, problems need to be well-typed, otherwise the translation outlined in 
Section~\ref{ss:erasure} might not be sound. 
Type checking in DTF thus takes on a larger role than in other logics in the TPTP World.

While performing the usual type checking procedure in DHOL, obligations of the form 
\(a\  t_1\ \cdots\  t_n \equiv a\  u_1\cdots\  u_n\), are generated.
These establish equality of the dependent base types applied to arguments
\(t_1\cdots t_n, u_1\cdots u_n\) of appropriate types. 
The type equality holds if all pairs \(t_i, u_i\) are equal, which depends on the available axioms.
This can create interesting situations where a problem must include axioms that are not necessary 
for proving the conjecture itself, but are necessary for type checking it.
The common example of fixed-length lists is one such example: 
the statement of the associativity of \texttt{append} is well-typed only if addition on $\nat$ is 
associative, and thus requires including the defining equations of addition.
To prove the problem only the defining equations of appending lists are needed.

The undecidability of type checking can lead to compromises. 
One such compromise is ``shallow type checking''.
When a problem file is shallowly checked, only the simply typed skeleton of the problem is
considered, i.e., term arguments to types as well as dependent functions are ignored. 
This collapses to type checking as is done on non-dependently typed problems, and is decidable.
This form of type checking is sufficient to catch many careless mistakes in the formulation of 
problems, and provides a basic check of issues often found in human-written DHOL problems. 
Examples are: mismatches in the number of arguments of a base type or function, and egregious 
type mismatches. 
Shallow type checking provides a valuable sanity check for users, especially considering the 
complexity that problems in DHOL forms can reach.

\subsection{Semantics}

As for HOL, there are two kinds of semantics for DHOL: standard models are intuitive and are the 
ones that are usually used; non-standard (Henkin) models are a generalization that is needed for 
completeness.
A full account is given in the forthcoming \cite{dhol_models}, which is summarized below.
The rules of DHOL, as given by Rothgang et al., already define which formulae are theorems.

\paragraph{Standard Models.}
Given a theory $T$, a standard model $M\in\sem T{}{}$ is a tuple providing an interpretation for 
every declaration in $T$.
Similarly, given a context $\Gamma$, an assignment $\alpha\in\sem\Gamma M{}$ for $\Gamma$ is a 
tuple providing an interpretation for every declaration in $\Gamma$.
These induce the interpretation function $\sem -M\alpha$ (with $\alpha$ omitted if the context 
is empty), which is defined inductively for all the syntax.
In particular, the possible components of a model are defined by induction on declarations:
\begin{compactitem}
\item For a type symbol with arguments $\Gamma=x_1:A_1,\ldots,x_n:A_n$, a function 
      $\sem\Gamma M{}\to\Set$
\item For a term symbol $c:A$, a value from $\sem A M{}$
\item For an axiom $\ass F$, a unique choice $\checkmark$ if $\sem M F{}=1$, and no choice 
      otherwise
\end{compactitem}
For the components of an assignment:
\begin{compactitem}
\item For a term variable $x:A$, a value from $\sem A M\alpha$
\item For an assumption $\ass F$, a unique choice $\checkmark$ if $\sem M F\alpha=1$, and no 
      choice otherwise
\end{compactitem}
For types and terms, the model is defined by induction in the usual way, in particular
\begin{compactitem}
\item $\sem \bool M\alpha=\{0,1\}$
\item $\sem{\dTo xAB}M\alpha$ is the set of functions $f$ mapping every $u\in\sem A M\alpha$ to 
      some $f(u)\in\sem BM{\alpha^u}$ where $\alpha^u$ extends $\alpha$ with the value $u$ for $x$
\end{compactitem}

\paragraph{General Models.}
The definition of general models generalizes the Henkin models from HOL by applying methods from 
categorical models of type theory.
First, akin to assignments for $\Gamma$, substitutions $\gamma:\Gamma\to\Delta$ as lists 
of terms or $\checkmark$ by induction on $\Gamma$ are defined:
\begin{compactitem}
\item For a term variable $x:A$, a term of type $\Delta\vdash A[\gamma]$
\item For an assumption $\ass F$, the unique choice $\checkmark$ if $\Delta\vdash F[\gamma]$, 
      and no choice otherwise
\end{compactitem}
Equality of contexts and substitutions is defined by applying the existing equality judgments for 
types and terms component-wise.
For every theory $T$, this yields the syntactic category $\gsem T$ of $T$-contexts and 
substitutions.
A general model is then defined as any pushout-preserving contravariant functor 
$\Phi:\gsem T\to \Set$.
From such a $\Phi$, an interpretation function is extracted using $\Phi(x:A)$ as the interpretation
of the type $A$ and $\Phi(t)$ as the interpretation of the term $t:A$ (seen as a substitution 
$x:A\to\bullet$).
These general models must further satisfy $\Phi(\bool)=\{0,1\}$, and $\Phi\models F$ is defined 
as $\Phi(F)=1$.
Here the pushout-preservation essentially corresponds to the preservation of substitution, i.e., 
interpretation and substitution commute.
The lack of any preservation of exponentials allows for a \emph{non-compositional} interpretation 
of function types.
This approach can be seen as a generalization of Henkin models, which also preserve substitution 
but do not need to interpret function types compositionally.
Contrary to Henkin models, the interpretation of $\lambda$ and application terms can also be non-compositional in these general models as long as substitution is preserved.

\paragraph{Models for Polymorphic DHOL.}
As mentioned above, a rank-1 polymorphic variant of DHOL is being developed in parallel work.
It is straightforward to extend standard models to polymorphic DHOL.
The syntax of binding a type variable corresponds to abstracting over an arbitrary set on the 
semantic side.
In particular, the interpretation of a polymorphic term/type symbol with $n$ type variables takes 
$n$ sets as arguments.
Polymorphic axioms correspond to universal quantification over sets.
The definition of syntactic category and general models is expected to carry over to polymorphic 
DHOL as well.
This has not been investigated in detail.

\section{Problem Dataset}
\label{s:probs}

Over 100 problems in DTF format have been collected for addition to the TPTP problem library.
Their classification is presented in Table~\ref{tab:cat} and discussed here (with
36 problems just for testing DHOL prover features omitted).
The number of problems in each class is given in the last column.
The problems concern several domains that can benefit from dependent types.
While \cite{RRB23} shows DHOL to be sound and complete, the strength of the existing automation
for this foundation (discussed in Section~\ref{s:impl}) still needs to be improved. 
For this reason, some of the harder problems were broken down into simpler subproblems that
can be proven independently. 
Some list properties that require both induction and reasoning with dependent types are an 
instance of this.
For example, the fact that list append is associative, \texttt{ListAppAssoc}, is split into three 
subproblems, showing the particular induction scheme, the proof of the base case, and the step 
case. 
These three subproblems are easier to prove than their combined version, which is also included.
Some problems benefit from intermediate lemmas, e.g. the instantiation of the inductive step case. 
These are found in the ``Lemmas'' categories of Table~\ref{tab:cat}.

One of the simplest classes of examples are lists that depend on their length (also called 
vectors, for example in the Rocq library). 
As the list libraries of most interactive theorem provers are substantial, it is relatively easy 
to experiment with many properties of dependently typed lists.
Such properties include the aforementioned associativity of \texttt{append}, corollaries of this 
statement, or involution statements about the \texttt{reverse} function.
Some of these list examples are extended to their polymorphic generalizations, which are in the 
``Polymorphic'' categories.

The idea of expressing well-known but sometimes challenging properties extends to several
other algebraic data types, such as matrices that have fixed dimensions, and lists of lists.
Red-black trees are a well-known data structure for balanced trees where the invariant can be 
expressed using dependent types, and again several problems concerning this type are included. 
The \texttt{Fin} type present in several proof libraries has been manually recreated,
and some problems about these are in the ROCQ category of Table~\ref{tab:cat}.
The collection includes the five examples from category theory that were originally presented in 
\cite{RRB23}, slightly reformatted to match the TPTP syntax.
To make use of the choice operator~\cite{RBK24}, several problems about dependent higher-order 
Skolemization are included.
Choice is also used in a function definition with no fixed point, and conjectures establishing
this are presented in the ``no FP'' category. 
Finally, several simple tests to evaluate the ability of provers to perform native DHOL 
inferences are provided.

Some of the dependent HOL problems are more interesting from a proof perspective -- the deep
type checking is there only to make sure the problem is well-formed. 
For example, for all the dependent list problems, the type checking obligations are there mostly 
to make sure no incorrect calls are being made, but they are relatively straightforward to
discharge. 
It is the proof that requires more logical reasoning. 
Other problems, while relatively straightforward in terms of proving, are harder to type check. 
This is because it is possible to use dependent types to encode important properties and 
invariants in the type system.

\begin{table}[htb]
\begin{center}
\begin{tabular}{llr}
  Problem Type & Problem Category & Problem Count\\
  \hline
Monomorphic Complete\hspace{3mm} & Category theory      &  5 \\
                     & Choice basic         & 11 \\
                     & Choice list          &  3 \\
                     & Choice no fixed point& 10 \\
                     & List app assoc       &  3 \\
                     & List app assoc corollary  &  1 \\
                     & List app nil         &  4 \\
                     & List of lists        &  1 \\
                     & List reversal involution  &  1 \\
                     & List reversal inv lemma   &  3 \\
                     & Matrices             &  5 \\
                     & ROCQ                 &  3 \\
\hline
Monomorphic Lemmas   & Choice no fixed point     & 10 \\
                     & List app assoc       &  5 \\
                     & List app assoc corollary    &  5 \\
                     & List reversal involution  &  5 \\
                     & List reversal inv lemma   & 11 \\
\hline
Polymorphic Complete & List app assoc poly  &  3 \\
                     & List app nil poly    &  4 \\
                     & List reversal involution poly\hspace{3mm} &  1 \\
                     & Red-black tree       &  3 \\
\hline
Polymorphic Lemmas   & List app assoc poly  & 14 \\
                     & List reversal involution poly & 13 \\
                     & Red-black tree       &  9 \\
\hline
\end{tabular}
\end{center}
\caption{The categories of the DTF problems.}
\label{tab:cat}
\end{table}

\section{Tools}\label{s:impl}

This section discusses the tools capable of processing problems in DTF format.

\subsection{The Logic Embedding Tool}

The Leo-III~\cite{SB21} prover includes the \emph{Logic Embedding Tool}, which has been extended 
to support polymorphic DTF.
The tool implements the erasure presented in Section~\ref{ss:erasure}, and incorporates 
the polymorphic extension.
The tool can generate both the type checking obligations and the translated problem separately.
This makes it possible to translate DTF problems into THF problems (that do not have dependent
types).
The embedding tool is available as NTFLET in SystemB4TPTP~\footnote{%
\href{https://tptp.org/cgi-bin/SystemB4TPTP}{tptp.org/cgi-bin/SystemB4TPTP}}.
The embedding tool enables the use of existing higher-order ATP systems for solving DTF problems,
by pipelining the output from NTFLET to a THF ATP system of the user's choosing.
This has been implemented as the DT2H2X ATP systems, available in SystemOnTPTP~\footnote{%
\href{https://tptp.org/cgi-bin/SystemOnTPTP}{tptp.org/cgi-bin/SystemOnTPTP}}.

\subsection{DLash}

The Lash prover~\cite{BK22} is a partial reimplementation of the tableaux calculus of 
Satallax~\cite{B12}, using a central term representation with perfect sharing. 
This design facilitated the implementation of the DLash extension of Lash, which handles 
DTF~\cite{NBK24}. 
In addition to the erasure implementation, DLash can process monomorphic dependently typed 
higher-order logic with choice. 
As with the Logic Embedding Tool, type checking and proving can be requested separately. 
DLash, like Satallax, includes a strategy language used to build so-called modes.
The current version includes 36 dedicated modes for dependent types, tailored to specific 
problem types.
DLash is available in SystemOnTPTP~\footnote{%
\href{https://tptp.org/cgi-bin/SystemOnTPTP}{tptp.org/cgi-bin/SystemOnTPTP}}.

\subsection{MMT}

MMT~\cite{rabe:recon:17} is a logical framework designed to formalize and manage large collections 
of interconnected formal systems and their libraries, using modular theory graphs.
A particular application of MMT is rapid prototyping~\cite{MR:prototyping:19}, and it was the 
tool originally used to develop and prototype DHOL.
The MMT/DHOL implementation offers reconstruction of omitted types and implicit arguments as well 
as parsing against user-defined notations.
It can be used to interactively author and type check DHOL problems and export them in TPTP format.
It uses the PER translation, and calls the Leo-III prover to discharge the resulting proof 
obligations.
MMT is mostly useful for developing formalizations, rather than proving TPTP conjectures.
Therefore, it does not provide a TPTP import at this point, but provides additional evidence of 
the well-typedness DTF problems.

\subsection{TPTP Systems}

As discussed in Section~\ref{ss:tptp}, TPTP includes several generic tools capable of processing 
problems and solutions.
For DTF problems:
\begin{itemize}
\item \texttt{TPTP4X} pretty-prints DTF problems and solutions, and offers various 
      transformations/augmentations of problems.
\item \texttt{BNFParser} produces the abstract syntax tree from parsing a DTF problem.
\item \texttt{Leo-III-STC} validates the syntax and types of DTF problems.
\item \texttt{ProblemStats} outputs various syntactic measures for problems.
\end{itemize}
All these tools are available in SystemB4TPTP~\footnote{%
\href{https://tptp.org/cgi-bin/SystemB4TPTP}{tptp.org/cgi-bin/SystemB4TPTP}}.
For DTF proofs:
\begin{itemize}
\item \texttt{ProofStats} outputs various syntactic measures for DAG-structured proofs.
\item \texttt{IDV} provides interactive viewing of proofs from DTF problems.
\end{itemize}
All these tools are available in SystemOnTSTP~\footnote{%
\href{https://tptp.org/cgi-bin/SystemOnTSTP}{tptp.org/cgi-bin/SystemOnTSTP}}.

\section{Conclusion}\label{s:conc}

This paper has described DTF, the dependently typed higher-order form of the TPTP language. 
It responds to the growing interest in dependently typed automated reasoning as exemplified by the number of TPTP
problems and tools that have cropped up in the short time since DHOL was first described. 
It can be seen as pushing the boundary of automated theorem proving towards language features that have previously been found only in interactive provers.

DHOL problems sometimes used differing standards, which defeated the uniformity advantage that 
the TPTP language provides.
This work unifies them, and provides over a 100 problems from different domains, benefiting
from the use of dependent types.
We hope that the availability of dependent types in the TPTP will stimulate research into
dependently typed automated theorem proving, by making it easier to exchange and compare results. 
Extending existing systems with support for DTF, and improving the performance of the systems that
already exist, will be important next steps.
In particular, the extension of superposition-based theorem proving to dependent types is a
tantalizing goal.

\subsubsection*{Acknowledgements:}
The authors thank Johannes Niederhauser and Colin Rothgang for granting access to their DHOL
 problems that are in the problem dataset. 
This work was supported by the ERC PoC grant no. 101156734 ``FormalWeb3''.

\bibliographystyle{splncs04}
\bibliography{references}

\end{document}